\documentclass{ws-ijprai}

\usepackage{graphicx}
\usepackage{amsmath}
\usepackage{amssymb}
\usepackage{tabularx}
\usepackage{graphicx}
\usepackage{subcaption}
\usepackage{times}
\usepackage{multirow}
\usepackage{multicol}
\usepackage{cite}
\usepackage{booktabs}

\begin{document}

\markboth{S.-E. Weng, S.-G. Miaou, R. Christanto}{A Lightweight Low-Light Image Enhancement Network via Channel Prior and Gamma Correction}

%
\catchline{}{}{}{}{}
%

\title{A Lightweight Low-Light Image Enhancement Network \\via Channel Prior and Gamma Correction}

\author{Shyang-En Weng}

\address{Department of Electronic Engineering,\\
Chung Yuan Christian University, \\Taoyuan, Taiwan\\
\email{shyangen104@gmail.com}
}
\author{Shaou-Gang Miaou}

\address{Department of Electronic Engineering,\\
Chung Yuan Christian University, \\Taoyuan, Taiwan\\
\email{miaou@cycu.edu}
}
\author{Ricky Christanto}

\address{Department of Electronic Engineering,\\
Chung Yuan Christian University, \\Taoyuan, Taiwan\\
\email{richrist81@gmail.com}
}

\maketitle


\begin{abstract}
Human vision relies heavily on available ambient light to perceive objects. Low-light scenes pose two distinct challenges: information loss due to insufficient illumination and undesirable brightness shifts. Low-light image enhancement (LLIE) refers to image enhancement technology tailored to handle this scenario. We introduce CPGA-Net, an innovative LLIE network that combines dark/bright channel priors and gamma correction via deep learning and integrates features inspired by the Atmospheric Scattering Model and the Retinex Theory. This approach combines the use of traditional and deep learning methodologies, designed within a simple yet efficient architectural framework that focuses on essential feature extraction. The resulting CPGA-Net is a lightweight network with only 0.025 million parameters and 0.030 seconds for inference time, yet it achieves superior performance over existing LLIE methods on both objective and subjective evaluation criteria. Furthermore, we utilized knowledge distillation with explainable factors and proposed an efficient version that achieves 0.018 million parameters and 0.006 seconds for inference time. The proposed approaches inject new solution ideas into LLIE, providing practical applications in challenging low-light scenarios.
\end{abstract}


\section{Introduction}

Human vision relies on the presence of light. The color and other visual characteristics of an object can be obtained through the reflected light irradiated on the object. Low-light environments have become an ongoing challenge for many applications, such as security surveillance, medical imaging, and autonomous driving. Since low-light images are usually characterized by low contrast, high noise, and a lack of details due to insufficient illumination, they may bring great difficulties to subsequent image analysis tasks. Low-Light Image Enhancement (LLIE) is an image processing technology that improves the characteristics of such images by restoring lost details, increasing contrast, and reducing noise, making them easier to clearly identify and analyze.

LLIE can be broadly categorized into two main approaches: histogram equalization and Retinex. Histogram equalization (HE) \cite{gonzales2017digital} is a fundamental technique for low-light image enhancement that redistributes grayscale intensity values in an image's histogram to enhance global contrast based on the original brightness. It provides both global and local changes, with global equalization being simple yet sometimes causing over-enhancement, information loss, and noise. Local HE enhances the global HE approach by preserving image details. It achieves this by dividing the image into regions, defining neighborhoods, traversing pixels, and applying continuous histogram equalization within these regions. This method helps retain fine-grained details while enhancing overall contrast. Adaptive HE (AHE) \cite{gonzales2017digital} refines this with local histograms, while Contrast Limited Adaptive HE (CLAHE) \cite{Pizer_1987_CVGIP} mitigates noise amplification. However, it has some limitations, such as information loss in over-enhanced regions and artifacts in enhanced images.

In contrast, Retinex \cite{land1977retinex} decomposes images into reflectance and illumination components, enhances reflectance to improve quality, and provides Single Scale Retinex (SSR) \cite{jobson1997properties} and Multi-Scale Retinex (MSR) \cite{Rahman_1996_ICIP} techniques. MSRCR \cite{jobson2002multiscale}  refines the MSR technique with color restoration, while MSRCP \cite{Petro_2014_IPOL} adds color balance, making it suitable for scenes with colored illumination. Retinex excels at preserving image detail and managing complex lighting conditions, but it requires massive computational resources, extensive training data, and complex algorithms to operate effectively.

Nowadays, with the development and application of deep learning, the huge potential of the Retinex algorithm has been greatly enhanced. Replacing or implementing rigorously defined image features with neural networks can achieve considerably higher-quality image enhancement capabilities. Deep learning has shown great potential in LLIE, including Retinex \cite{Wei_2018_BMVC, zhang2019kindling} and generalized network design \cite{Zamir_2020_ECCV, Wang_2022_AAAI}. However, most existing deep learning models require large amounts of high-performance computing resources, making them less suitable for use in embedded systems or low-power devices.

The pursuit of lightweight deep learning models tailored for edge computing remains a formidable challenge. There is a growing demand for these models that can be efficiently and accurately enhanced while being optimized for low memory and computational resources. This optimization makes them ideal for real-time applications and resource-efficient devices. Numerous strategies have emerged to develop such lightweight deep learning models, including network compression \cite{Park_2018_IEEEAccess}, knowledge distillation \cite{Li_2023_Neurocomputing}, and architectural design \cite{Wang_2020_TIP, lamba2021restoring, Ma_2022_CVPR, liu2021ruas}. Also, it is a good way to enhance the entire model with pertinent features, e.g. \cite{Zhang_2021_arXiv} and \cite{Cui_2022_BMVC}. Therefore, we utilize pertinent features directly with the corresponding system architecture to minimize the computational effort, which presents another promising approach to achieving these goals. 

Our key contributions include:
\begin{itemize}
    \item Integration of Traditional Methods: We proposed CPGA-Net (Channel Prior and Gamma Estimation Network), a lightweight convolutional approach that efficiently combines multiple traditional enhancement elements, including dark and bright channel priors (DCP/BCP), luminance (Y) from YCbCr, and gamma correction with Intersection-Aware Adaptive Fusion Module (IAAF).
    \item Effectiveness: We evaluated the proposed method with the LLIE benchmark, and it achieves state-of-the-art performance with a relatively low number of parameters and computational cost. Also, it converges fast with only 50 epochs, and 30 epochs for transfer learning on similar datasets to reach the best performance.
    \item Interpretability: We present comprehensive experimental results, conduct an ablation study, and demonstrate the interpretability of CPGA-Net for LLIE, elucidating the functionality of each component and feature. Furthermore, we leverage this characteristic in conjunction with knowledge distillation, successfully compressing the computational cost while maintaining high performance.
\end{itemize}

\section{Related Works}
In the ever-evolving field of LLIE, a range of methods and techniques have emerged to address the challenging task of enhancing images captured under suboptimal lighting conditions. These methods cover a variety of approaches. We will delve into previous LLIE methods, explore the relationship between LLIE and dehazing, and investigate pertinent features typically employed in LLIE.
Retinex Theory \cite{land1977retinex} has been used to develop a number of LLIE techniques. The core idea behind this theory is that the color of an object is determined by the relative intensities of the light it reflects, rather than the absolute amount of light. In mathematical terms, for each pixel location at $(i, j)$, it can be expressed as Eq. \eqref{eq:Retinex}:
\begin{equation}
I(i, j) = L(i, j) \cdot R(i, j) \label{eq:Retinex}
\end{equation}
where $I$ is the perceived color, $L$ is the illumination, and $R$ is the reflectance of the object. The LLIE techniques based on this theory first decompose an image into its constituent brightness components. They then enhance each brightness component independently. This helps improve the overall brightness and contrast of the image while preserving its natural appearance.

Traditional Retinex methods for LLIE often rely on handcrafted features and rules. The most famous method using Retinex is LIME \cite{guo2016lime}, which follows the principle of Retinex and mainly estimates illuminance for restoration. It formulates this estimation task as an optimization problem, partitioning the horizontal and vertical components of illumination into three variables and determining them using the Augmented Lagrangian Multiplier. This approach effectively restores the original image based on its gradients and structural priors. Other methods utilizing Retinex include \cite{jobson1997properties, Rahman_1996_ICIP, jobson2002multiscale, Petro_2014_IPOL, Zhang_2019_CGF, fu2016weighted}. Retinex is also widely employed in deep learning for LLIE, typically following a two-stage training process involving decomposition and adjustment. In the decomposition stage, the image is separated into reflectance and illumination based on Retinex Theory and certain image characteristics. Subsequently, during the adjustment phase, illumination is enhanced, and the reflectance is de-noised independently and then merged together. This process effectively enhances the overall brightness and contrast of the image while maintaining its natural appearance, as demonstrated in \cite{Wei_2018_BMVC, zhang2019kindling}.

The Atmospheric Scattering Model (ATSM) \cite{McCartney_1976, Narasimhan_2000_CVPR, Narasimhan_2002_IJCV} plays a crucial role in LLIE by describing how atmospheric particles scatter light and aid in image dehazing, as shown in Eq. \eqref{eq:ATSM} :
\begin{equation}
    I(x) = J(x)t(x) + A(1 - t(x)) \label{eq:ATSM}
\end{equation}
where $I(x)$ represents the hazy image, $J(x)$ corresponds to the haze-free image, $A$ denotes the airlight value, and $t(x)$ signifies the transmittance matrix.

ATSM helps estimate critical parameters such as scene transmission and airlight, which are essential for dehazing. Furthermore, Dark Channel Prior (DCP) \cite{he2010single} is a heuristic method that shows that the dark channel of a hazy image is generally quite dark. This occurs because it represents the minimum of all RGB channels, with haze particles scattering more light in the brighter channels. \cite{dong2010fast} noted the similarity between inverted low-light images and haze images and incorporated the concepts of Retinex Theory into ATSM. Thus, it is important to recognize the close relationship between image dehazing and low-light image enhancement. Several works have explored the use of ATSM, such as \cite{dong2010fast, Zhang_2012_ICPR, li2015low, Shi_2018_EURASIP, jiang2013night}.

For the advanced methods of LLIE, most are based on deep learning and can be divided into physical models and non-physical models. Physical models, such as \cite{Wei_2018_BMVC, zhang2019kindling}, usually follow the Retinex Theory or ATSM; non-physical models, such as \cite{Zamir_2020_ECCV, Wang_2022_AAAI, lamba2021restoring, Ma_2022_CVPR}, are implemented using various machine learning algorithms or specific deep learning architectures, and some are also suitable for general degradation tasks.

Drawing on ideas from numerous previous works, we know that some pertinent features can effectively represent the essence of low-light images, even in the presence of information loss and noise. This may be a possible way to enhance deep learning methods through prior knowledge. For instance, as shown by \cite{Shi_2018_EURASIP, Wang_2013_SignalProcessing}, Bright Channel Prior (BCP), which is determined by the maximum values among all RGB channels, can preserve the most prominent features of the image. On the other hand, the Dark Channel Prior has proven to be less effective in LLIE than in image dehazing since the lowest values in low-light images represent their darkness and are difficult to manipulate. Interestingly, however, \cite{Shi_2018_EURASIP} indicated that the dark channel can retain the missing structures found in the bright channel, suggesting that it might serve as a valuable complement to help correct estimated transmission and ultimately enhance overall performance by combining both bright and dark channels. HEP \cite{Zhang_2021_arXiv} used the feature maps of the HE-enhanced image as priors to guiding the restoration process and implemented unsupervised LLIE through noise disentanglement. IAT \cite{Cui_2022_BMVC} employed a global branch with color correction and gamma correction, which significantly improves the performance and demonstrates effectiveness in LLIE, exposure correction, and object detection tasks. 

In our previous work \cite{Weng_2023_IS3C}, we found that using super-resolution (SR) methods to reduce image resolution can successfully reduce the computational load and maintain image quality in dehazing through joint training of dehazing and SR networks. Furthermore, we also know that the combined model with bicubic interpolation surprisingly produces better results and demonstrates the potential of using conventional algorithms as part of a deep learning approach. Recently, some research has started to explore new tasks that combine SR and LLIE \cite{aakerberg2021rellisur}. For example, \cite{Rasheed_2022_Neurocomputing} combined the specialized structure for SR by learning enhanced and dark features in low-resolution space and then using sub-pixel layers to upsample the features into the image. However, these methods, although powerful, require larger and more specific computations to handle both tasks simultaneously. DGF (Deep Guided Filter) provides a solution to reduce image resolution and maintain image quality by integrating guided filters into deep learning \cite{Wu_2018_CVPR}, making the model faster and achieving state-of-the-art performance. In our work, we exploit it as an extension module to reduce the computational load.

In summary, extensive research in the field of LLIE has produced a wealth of techniques that provide valuable insights into capturing the essence of low-light images, often under challenging conditions with noise and information loss. Leveraging rich prior knowledge and combining it with the potential of deep learning marks substantial advancement in our pursuit of efficient and effective LLIE methods. Our proposed CPGA-Net incorporates many feature extraction and representation concepts to provide better image enhancement performance and may become a promising approach for the continued development of this field.

\section{Proposed Method}
\subsection{Architecture of CPGA-Net} \label{section: arch}
There is a close relationship between the Retinex Theory ~\cite{land1977retinex} and the Atmospheric Scattering Model \cite{McCartney_1976, Narasimhan_2000_CVPR, Narasimhan_2002_IJCV}. This relationship suggests that inverted low-light images have similar characteristics to hazy images. Based on LIME \cite{guo2016lime}, $I(x)$ and $J(x)$ in the Atmospheric Scattering Model are replaced with $[1-L(x)]$ and $[1-R(x)]$, respectively, leading to Eq. \eqref{eq:RevATSM}:
\begin{equation}
    [1 - L(x)] = [1 - R(x)] \cdot t(x) + A(x) \cdot [1 - t(x)] \label{eq:RevATSM}
\end{equation}
where $L$ is the low-light image, $1-L$ is the inverted low-light image, $R$ is the normal-light source image of interest, and $1-R$ is the inverted normal-light source image. Thus, we can get $R(x)$ through Eq. \eqref{eq:LLformula}:
\begin{equation}
    R(x) = \frac{L(x) - \widetilde{A}(x)}{t(x)} + \widetilde{A}(x) \label{eq:LLformula}
\end{equation}
where $\widetilde{A}(x) = 1 - A(x)$. Eq.~\eqref{eq:LLformula} is incorporated into the proposed neural network architecture, as shown in Fig.~1. In this study, we consider the dark channel and bright channel priors as the transmission term to estimate the balance between low-light images and $A(x)$ and perform gamma correction to adjust global influence. If $\widetilde{A}(x) = 0$, we can figure out that Eq. \eqref{eq:LLformula} will be the same expression as Eq.~\eqref{eq:Retinex}, which represents the Retinex Theory. If we look more closely at Eq. \eqref{eq:LLformula}, it is quite similar to the formula for image dehazing which can be written as Eq.~\eqref{eq:image_dehazing} by the atmospheric scattering model. 
\begin{equation}
    J(x) = \frac{I(x) - A(x)}{t(x)} + A(x) 
    \label{eq:image_dehazing}
\end{equation}
This characteristic shows the connection between LLIE and Image Dehazing, meaning that they are similar tasks to be solved.

\begin{figure*}[ht]
    \centering

    \includegraphics[width=\textwidth]{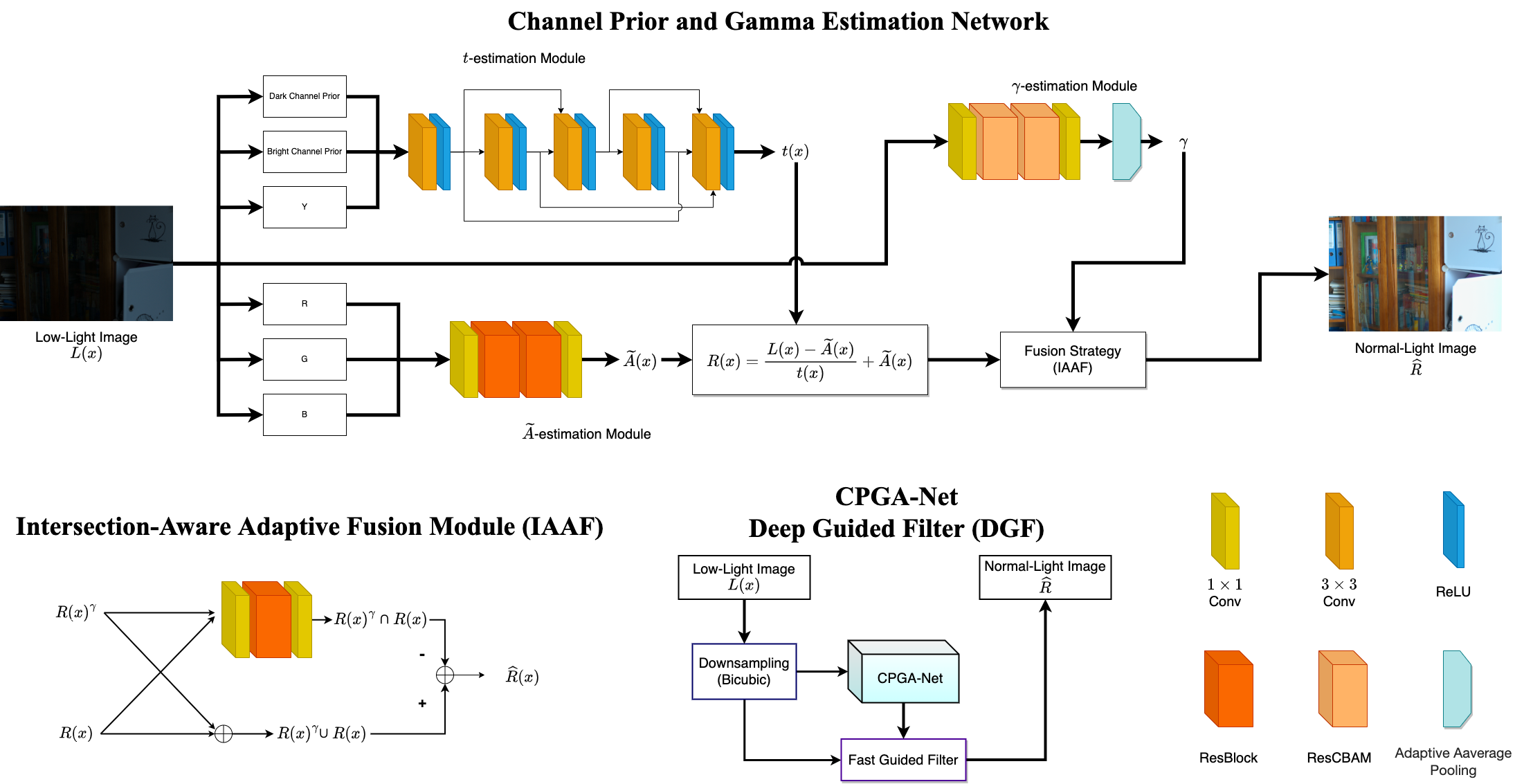}
    \caption{The architecture of CPGA-Net.}
    \label{fig:Arch}
    \hfill
\end{figure*}

Furthermore, we also propose the Intersection-Aware Adaptive Fusion Module (IAAF) to prevent divergence caused by the exponential term ($\gamma$) in gamma correction. Therefore, our LLIE method uses specialized features to improve performance and then reduce the computational load of the model, which we call CPGA-Net (Channel Prior and Gamma Estimation Network) to facilitate further discussion, as shown in Fig.~\ref{fig:Arch}.

We also integrated the Fast Guided Filter module into the original CPGA-Net while reducing the number of channels in the local branch to obtain the DGF version of CPGA-Net. In the DGF version, the input low-light image and its reduced-resolution counterpart become the inputs to the Fast Guided Filter module, and the output will be the enhancement result of the low-light image. This DGF version can provide a lower computational load and faster speed than the original CPGA-Net.

\subsection{$t$-estimation of CPGA-Net}
Estimating ``lightness'' is a multifaceted challenge in the field of image processing. The intricate interplay of factors such as varying lighting conditions, color balance, and the subjective nature of human vision is the essence behind this challenge. In the pursuit of LLIE, the significance of Dark Channel Prior (DCP) \cite{he2010single} and Bright Channel Prior (BCP) \cite{Wang_2013_SignalProcessing} becomes evident, as demonstrated in \cite{li2015low, jiang2013night, Shi_2018_EURASIP, Wang_2013_SignalProcessing, lee2020unsupervised, Tao_2017_ICIP}. The two priors can be represented as Eqs. \eqref{eq:BCP} and \eqref{eq:DCP}:
\begin{equation}
    I^{\text{bright}}(x) = \max_{c \in \{r, g, b\}} \left( \max_{y \in \Omega(x)} \left( I^c(y) \right) \right) \label{eq:BCP} 
\end{equation}
\begin{equation}
    I^{\text{dark}}(x) = \min_{c \in \{r, g, b\}} \left( \min_{y \in \Omega(x)} \left( I^c(y) \right) \right) \label{eq:DCP} 
\end{equation}
where \(I^c\) is a color channel of the input image \(I\) and \(\Omega(x)\) represents the local patch centered at position \(x\). These priors exploit the unique characteristics of dark and bright pixels within images and provide the essential cues to enhance low-light images while preserving key details. Accurate estimation of lightness is crucial to precisely controlling the effects of restoration processes and preventing the pitfalls of over-enhancement or unnatural defects.

\begin{figure}[b]
    \centering
    \includegraphics[width=0.7\columnwidth]{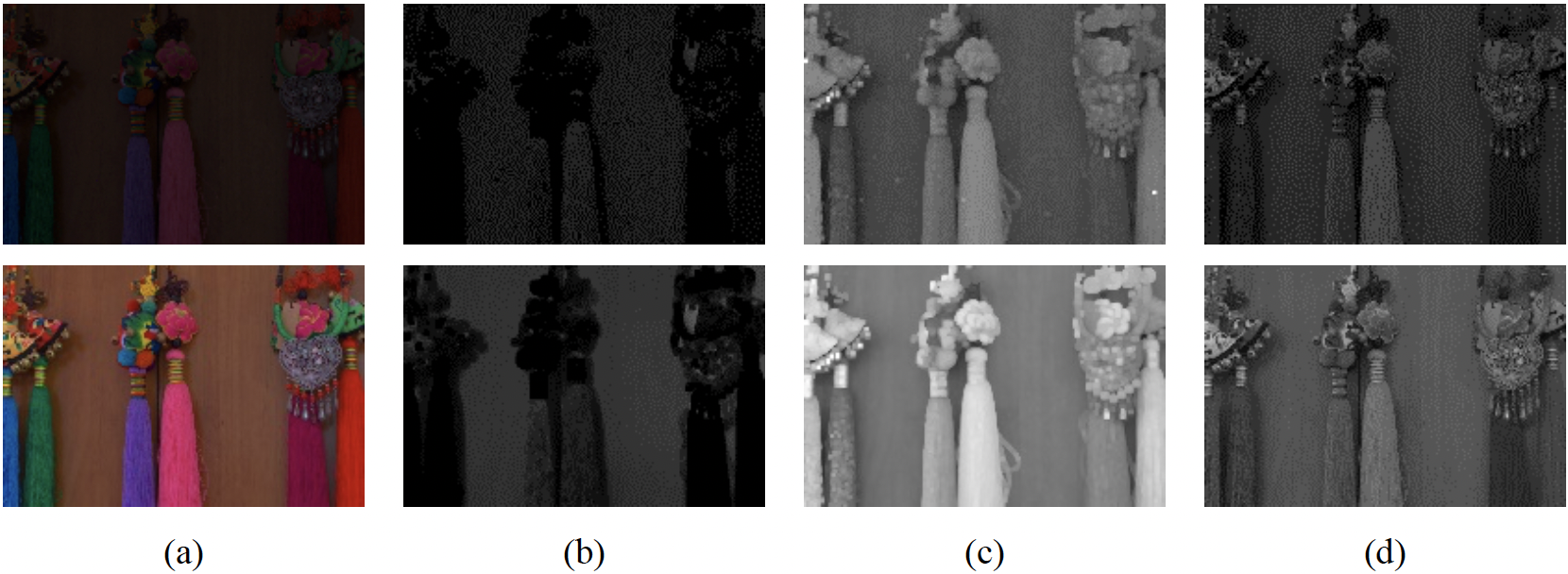}
    \caption{Examples of multiple channel priors. Top row: low-light images; bottom row: normal-light images. (a) Original image; (b) Dark channel prior; (c) Bright channel prior; (d) Luminance Y from YCbCr. For the visibility of observing channel priors of low-light images, we adjusted them with a 40\% brightness enhancement to make their features more apparent.}
    \label{fig:low_light_channel_prior}
\end{figure}

Notably, the bright and dark channel priors capture specific image attributes — illuminating regions affected by haze or low lighting and highlighting areas with atmospheric scattering or bright objects, respectively. Beyond these priors, a variety of methodologies exist for gauging lightness within conventional frameworks, each offering distinct insights into perceptual luminance. We extract channel priors directly without a local patch for the complete structure preserved for the learning of our neural network, as shown by Eqs. \eqref{eq:bright_channel} and \eqref{eq:dark_channel}:
\begin{equation}
    I^{\text{bright}}(x) = \max_{c \in \{r, g, b\}} \left( I^c(x) \right) \label{eq:bright_channel}
\end{equation}
\begin{equation}
    I^{\text{dark}}(x) = \min_{c \in \{r, g, b\}} \left( I^c(x) \right) \label{eq:dark_channel}
\end{equation}
Fig.~\ref{fig:low_light_channel_prior} shows the channel priors of low-light and normal-light images.

\subsection{$\widetilde{A}$-estimation of CPGA-Net}
The estimation of $\widetilde{A}$ plays a key role in the image reconstruction process using original RGB images as input. According to Eq. (4), it serves to identify and rectify the imperfections and flaws arising from image quality degradation. The limitations of conventional histogram equalization become apparent when processing low-light images, as they inherently suffer from information loss due to their darker nature. Contrast adjustment alone proves inadequate to compensate for this loss, necessitating the use of restoration techniques, as exemplified by Retinex-based algorithms such as MSR~\cite{Rahman_1996_ICIP} and LIME~\cite{guo2016lime}. 

Currently, advanced neural networks with deeper and more specialized architectures are well-suited to handle a wide range of image enhancement tasks, including those involving low-light conditions. Examples include the Swin Transformer structure employed in SwinIR \cite{liang2021swinir} and the multiple attention strategy featured in MIRNet \cite{Zamir_2020_ECCV}. Clearly, a robust feature extractor greatly affects the overall efficacy of a neural network. 

In our study, prioritizing computational efficiency, we adopt a simple architecture consisting of two layers of ResBlock for $\widetilde{A}$-estimation, demonstrating its effectiveness within our streamlined framework. Moreover, we conduct a comparative analysis involving the use of two CNN layers to simulate various scenarios and assess the impact of $\widetilde{A}$ on the enhancement process.

\subsection{Global Branch of CPGA-Net}
In the field of image processing, a variety of methods have emerged to use global features to improve image quality. Contrast stretching is one of them, and it is effective in various applications \cite{gonzales2017digital}. It works by linearly or non-linearly mapping the original range of pixel values to a new range. The process involves selecting the minimum and maximum pixel values from the original image and then stretching the pixel values in between to cover the entire range of possible values, including some mapping functions such as rubber band functions, gray‐level slicing, bit‐plane slicing, and more. These methods propose a new dynamic range to demonstrate and highlight the characteristics of specific grayscales.

When it comes to deep learning methods, feature extraction plays a crucial role in generating enhanced images from input images. The two well-known architectures frequently used are convolutional neural networks (CNNs) \cite{lecun1998gradient} and vision transformers (ViTs) \cite{dosovitskiy2020vit}. No matter which of them is used in image enhancement, they convert pixels into new values by mapping the image with a non-linear mapping function. Inspired by contrast stretching, our approach employs an exponential function to map images, steering non-linear mapping and contrast factors toward conditions resembling normal lighting. It is important to recognize that CNN and ViT functions primarily operate within the spatial domain and do not address, by themselves, the gamma factor adjustment problem. Also, we follow the implementation of IAT~\cite{Cui_2022_BMVC} and use global branches to improve performance; our approach introduces novel adaptations based on necessary features. In the context of our global branch, we deviate from the conventional path by modifying the gamma estimation process. The gamma correction is shown in Eq. (10):
\begin{equation}
    s = c r^\gamma \label{eq:processing_equation}
\end{equation}
where $r$ and $s$ represent the values of pixels before and after processing, respectively. The constant $c$ is set to 1.

Moreover, effectively controlling the local branch operating in the spatial domain and the global branch operating in the exponential domain presents a complex task given the convergence challenges during training. This complexity can complicate training goals and make the process prone to divergence. Taking these aspects into account, we introduce a strategic approach as Intersection-Aware Adaptive Fusion Module (IAAF). This involves quantifying the correspondence between the output of the local branch and the gamma-corrected output, preserving the distinctive features of the spatial and exponential domains while simultaneously exercising control over both branches. The idea is expressed as \break Eq. (11):
\begin{equation}
    \hat{R} = R \cup R^\gamma - (R \cap R^\gamma) \label{eq:fusion_strategy}
\end{equation}
In Eq. (11), the enhanced image $\hat{R}$ is the union of $R$ and $R^\gamma$ minus their intersection, estimating the similarity between two different domains and enhancing it directly by pixelwise addition, balancing the enhancement of the local branch and global branch. Compared with other fusion operations such as weighted calculations and direct concatenation, reconstructing images by considering the connections and differences between both tasks is a more intuitive and reasonable strategy.

By selectively removing redundant features, we maintain model convergence and improve performance. We implement this technique using ResCBAM~\cite{Woo_2018_ECCV}, which is \break CBAM (Convolutional Block Attention Module) integrated with ResBlock. CBAM introduces spatial and channel-attention mechanisms and demonstrates excellent performance in classification and object detection tasks, making it suitable for the purpose of our gamma estimation. In our ablation study, we will comprehensively evaluate network selection and IAAF modules.

\subsection{Training Strategy and Model Compression}
In this section, we will introduce two of the strategies we used in the CPGA-Net and its efficient version called CPGA-Net (DGF). 
In the training strategy of CPGA-Net, we opted for a combination of L1 loss and Perceptual loss. The utilization of L1 loss is common in image enhancement tasks. Perceptual loss~\cite{johnson2016perceptual} is a type of loss function commonly used in image restoration, style transfer, and image generation. It emphasizes the capture of high-level features and structures that closely resemble human perception. The loss function is expressed as Eq.~\eqref{eq:lossfuction}.

\begin{equation}
    L_{\text{enhance}} = \lambda_1 \cdot L_1 + \lambda_2 \cdot L_{\text{per}} = \lambda_1 \cdot \lVert \hat{Y} - Y^{GT} \rVert_1 + \lambda_2 \cdot \lVert \Psi(\hat{Y}) - \Psi(Y^{GT}) \rVert_2^2
    \label{eq:lossfuction}
\end{equation}
where $\hat{Y}$ is the output and $Y^{GT}$ is the ground truth; $\Psi$ represents the feature extractor of VGG16; $\lambda_1$ and $\lambda_2$ represent the loss weights, and they are empirically set to 1 and 0.01, respectively. It's important to note that our perceptual loss is a simplified version of the original formulation.

\begin{figure}[ht]
    \centering
    \includegraphics[width=\textwidth]{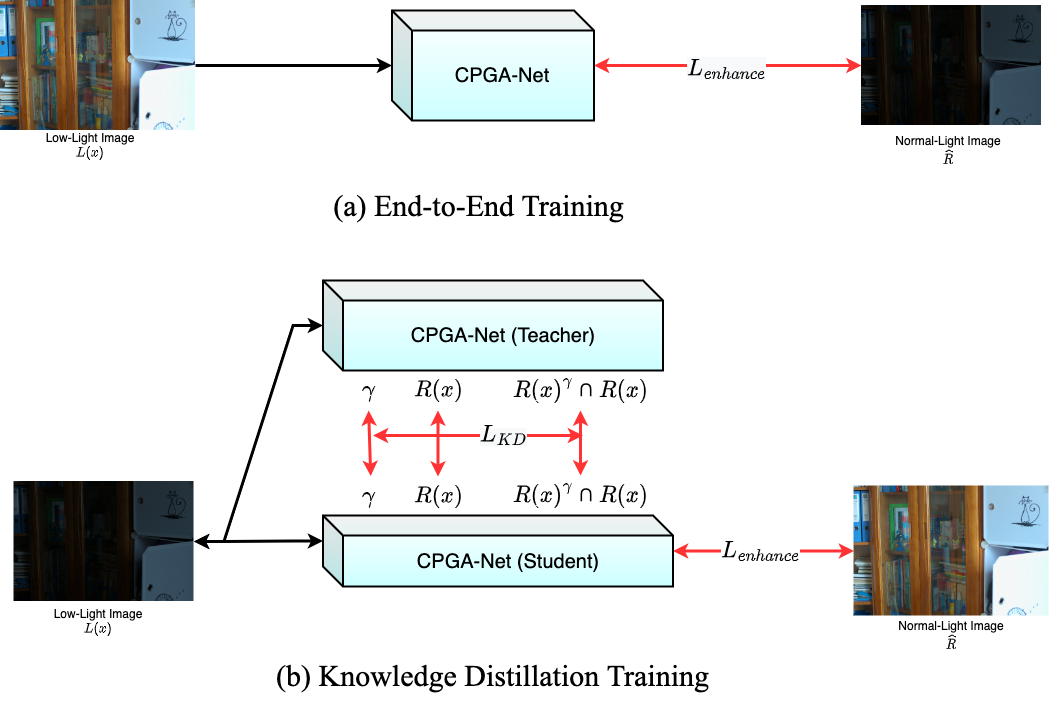}
    \caption{Training strategies of CPGA-Net and CPGA-Net (DGF).}
    \label{fig:training_strategy}
\end{figure}

In contrast to the typical supervised end-to-end training approach for image enhancement, CPGA-Net (DGF) adopts a distinctive strategy involving knowledge distillation based on our theoretical neural network model. In this knowledge distillation process, we select $R, \gamma, \text{and } R \cap R^\gamma$ as the crucial components. These choices aim to prompt the model to capture more detailed information, as illustrated in Fig.~\ref{fig:training_strategy}.

Unlike the common approach of mapping distillation maps into shallower layers for comparison between teacher and student models, we directly map them due to the unique characteristics of our theory-based model. Despite the inherent variability introduced by the ``Black box'' nature of training, our model's interpretability, supported by theoretical equations in section~\ref{section: arch} and visualizations in section~\ref{section: interpretability}, indicates that the selected components may indeed display similar features, as shown in Eq.~\eqref{eq:LLformula}.

To reinforce the resemblance between the teacher and student models, we supervise the three components — $R, \gamma, \text{and } R \cap R^\gamma$ — using the mean squared error (MSE), as indicated in Eq.~\eqref{eq:knowledge_distillation}. It is noteworthy that we do not supervise the output directly, as the output is constructed using these three components with numerical operators. Both the teacher and student models undergo pretraining before incorporating this technique.

\begin{equation}
    L_{\text{KD}} = \sum_{x \in \{\gamma, \widetilde{A}, R \cap R^{\gamma}\}} \lVert x_{\text{teacher}} - x_{\text{student}} \rVert_2^2 
    \label{eq:knowledge_distillation}
\end{equation}
Our knowledge distillation loss is defined by computing the Mean Squared Error (MSE) between the components $\gamma$, $\widetilde{A}$, $R \cap R^\gamma$ extracted from the student and teacher models. Hence, the loss function of CPGA-Net (DGF) is shown in Eq.~(14).

\begin{equation}
    L = \lambda_1 \cdot L_1 + \lambda_2 \cdot L_{\text{per}} + \lambda_3 \cdot L_{\text{KD}}
    \label{eq:dgf_lossfuction}
\end{equation}
where $\lambda_3$ is the loss weight of knowledge distillation loss; $\lambda_1, \lambda_2,\text{and } \lambda_3$ are empirically determined as 1, 0.01, and 0.1, respectively.

The training strategy for CPGA-Net (DGF) involves a multi-stage process. Initially, the CPGA-Net structure undergoes independent training with Eq.~\eqref{eq:lossfuction}. Subsequently, the first fine-tuning stage is conducted with Eq.~\eqref{eq:dgf_lossfuction}. Finally, a second stage of fine-tuning is performed independently with the CPGA-Net (DGF) structure with Eq.~\eqref{eq:lossfuction}. The differences between strategies will be analyzed in the ablation study to assess their respective contributions and impact on the overall model performance.

\section{Experiments Results}
\subsection{Datasets and Evaluation Metrics}
In our study, we trained and tested LLIE networks on two benchmark datasets: LOLv1 (LOw-Light dataset) \cite{Wei_2018_BMVC} and LOLv2 \cite{Wei_2018_BMVC}, respectively. These two datasets are designed specifically for LLIE. The LOLv1 dataset comprises a total of 500 images, divided into 485 for training and 15 for testing. On the other hand, the LOLv2 dataset consists of 1000 images, including both synthetic and real data. To assess the performance of LLIE networks on LOLv1, our training dataset includes the original 485 images along with an additional 1000 synthetic images, and the evaluation is carried out on its designated test set consisting of 15 images, following a procedure similar to that of RetinexNet \cite{Wei_2018_BMVC}. As for LOLv2, the training set is composed of real captured images, and the evaluation is performed on its test set, which contains 100 images. Also, we tested on real and unpaired datasets, including LIME \cite{guo2016lime}, NPE \cite{Wang_2013_TIP}, and MEF \cite{Ma_2015_TIP}, consisting of 10, 84, and 17 samples, respectively.

In the performance assessment of our LLIE method, we employ a comprehensive set of metrics to evaluate its effectiveness. These metrics include Peak Signal-to-Noise Ratio (PSNR), Structural Similarity Index Measure (SSIM), and Learned Perceptual Image Patch Similarity (LPIPS) \cite{Zhang_2018_CVPR}. For unpaired datasets where these full reference metrics are not available, we use the Naturalness Image Quality Evaluator (NIQE) for evaluation.

Efficiency is a critical consideration in image enhancement methods, especially for real-time applications. To measure efficiency, we consider two key factors: the number of parameters, which quantifies the model's size and reflects the number of learnable parameters, and FLOPs, which measure the computational workload of the model by quantifying the number of floating-point operations performed per second during inference. Lower FLOPs values suggest that the method can achieve results with fewer computational resources, making it suitable for real-time or resource-constrained applications. For FLOPs, we follow the implementation of \cite{guo2023survey} and \cite{hu2021two} with an image size of $3 \times 600 \times 400$, which is the same as the image size of the LOL dataset.

\subsection{Implementation Details}
The equipment used in this study includes a CPU (Intel Core i9-12900), 32 GB of RAM, and a GPU (Nvidia GeForce RTX 3090) for testing and evaluation. For the training of our deep neural network, the ADAM optimizer is used, the learning rate is set to $10^{-4}$, the epoch is set to 50, the batch size is set to 8, and L1 loss and perceptual loss are used as the loss functions. The training time per epoch is less than a minute.

For model initialization, we employ self-supervised training with 20 epochs, utilizing input images as targets to prime the model. This approach is exclusively applied to the train-from-scratch model. For LOLv2 and each fine-tuning stage of the training strategy for CPGA-Net (DGF), the networks were fine-tuned with 30 epochs and learning rate $10^{-5}$ using LOLv1 pretrained weights. CPGA-Net and CPGA-Net (DGF) employ different numbers of Resblock channels to improve performance and efficiency. In the regular version of CPGA-Net, we use 16 channels, while in CPGA-Net (DGF), we use 8 channels with a fast guided filter to reduce computational load and inference time.

\subsection{Results and Discussion}
\begin{figure*}[ht]
    \centering
    \includegraphics[width=\textwidth]{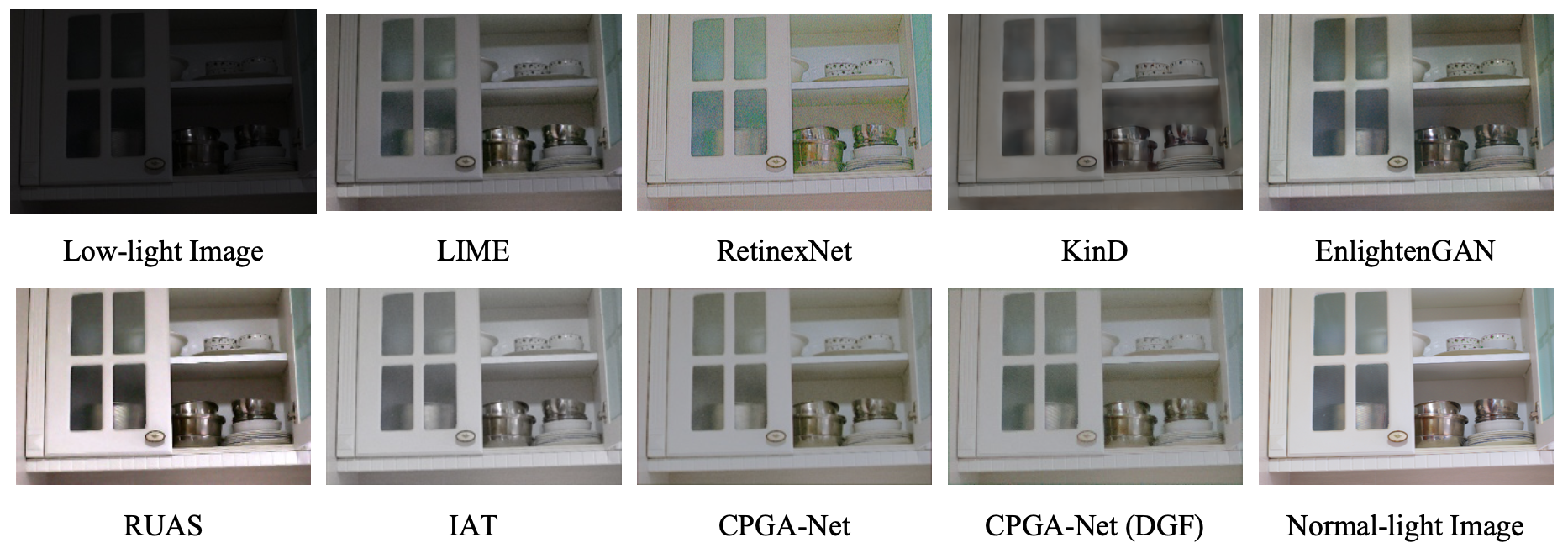}
    \caption{Visualization results of different methods on the LOLv1 dataset.
}
    \label{fig:v1}
\end{figure*}

\begin{figure*}[ht]
    \centering
    \includegraphics[width=\textwidth]{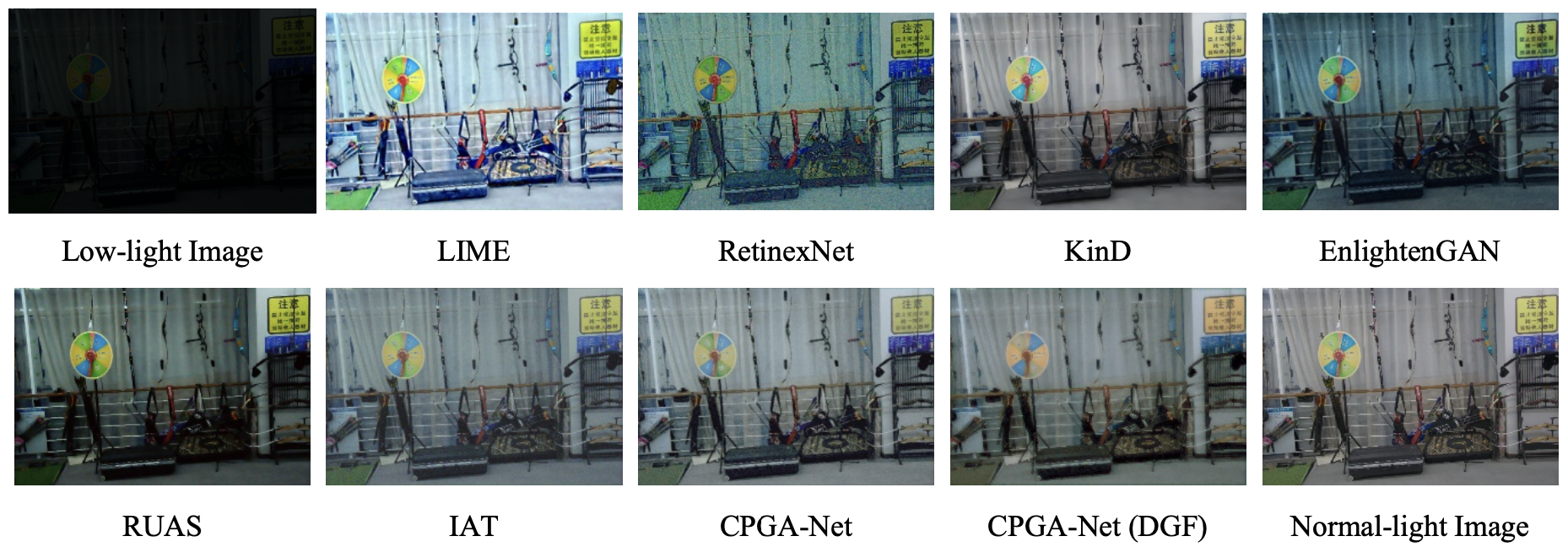}
    \caption{Visualization results of different methods on the LOLv2 dataset.
}
    \label{fig:v2}
    
\end{figure*}

\begin{figure}[ht]
    \centering
    \includegraphics[width=\textwidth]{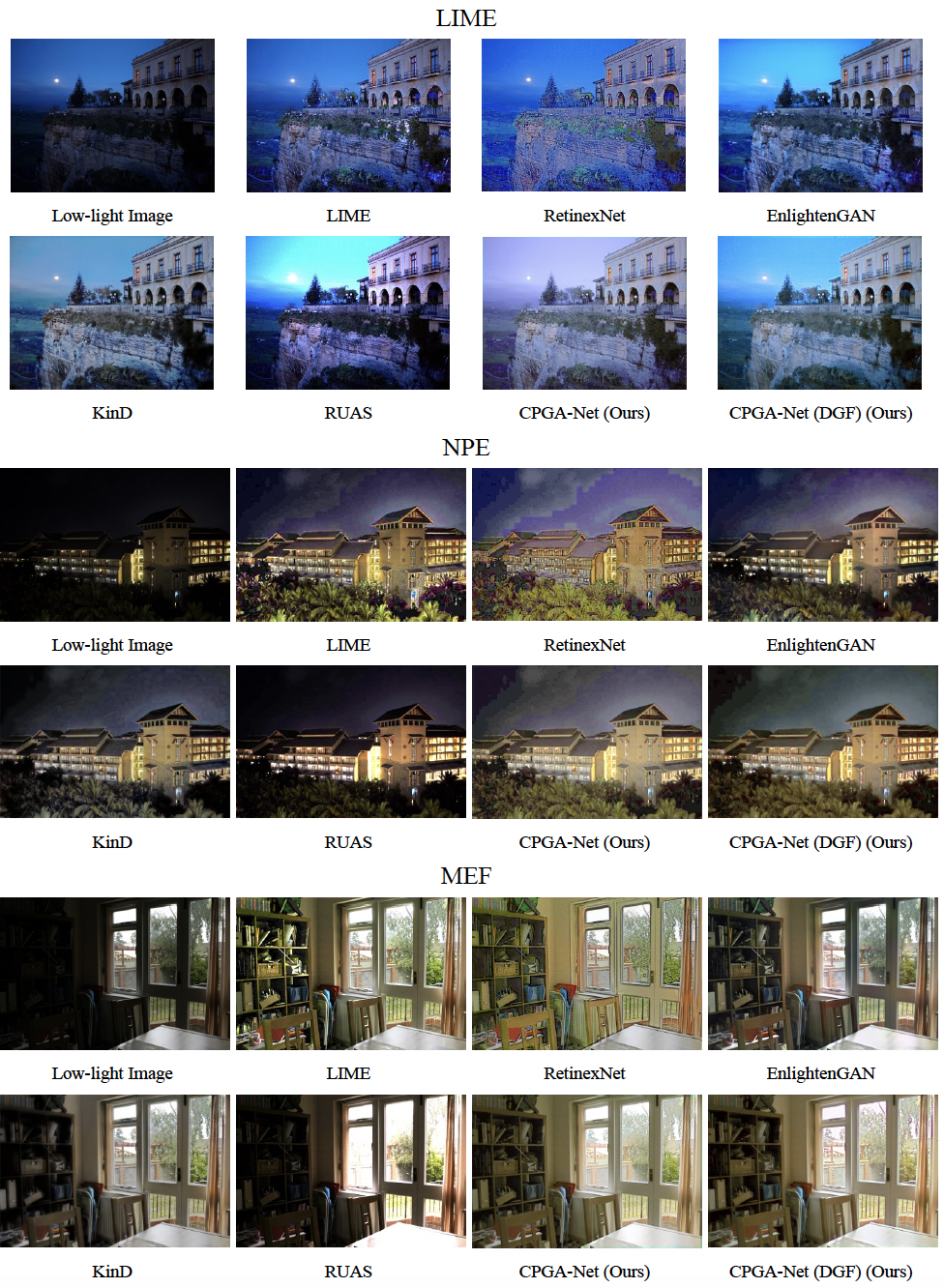}
    \caption{Visualization results of different methods on the unpaired datasets.}
    \label{fig:unpair}
\end{figure}

In our evaluation, we compared our method against state-of-the-art (SOTA) techniques as shown in Tables~\ref{tab:performance_metrics}, \ref{tab:efficiency_results}, and
\ref{tab:niqe_results} and Figs. \ref{fig:v1}, \ref{fig:v2}, and \ref{fig:unpair}. In Table \ref{tab:performance_metrics}, note that CPGA-Net achieves impressive results in terms of image quality metrics, ranking firmly in the second and third places. Notably, it achieves excellent performance on the LOLv1 and LOLv2 datasets while maintaining remarkably low numbers of parameters and FLOPs for inferencing, as shown in Table \ref{tab:efficiency_results}. This demonstrates the efficiency of our approach in achieving SOTA results. Also, it performs well using NIQE on unpaired datasets, as shown in Table \ref{tab:niqe_results} and Fig. \ref{fig:unpair}. Our approach is unique because it uses a simple but effective method to extract features. It aligns with the atmospheric scattering model and demonstrates that CPGA-Net, with a fast guided filter and slightly reduced size, can improve computational efficiency while maintaining image quality.

Compared with another lightweight and efficient structure, IAT \cite{Cui_2022_BMVC}, our method exhibits a significant gap in SSIM values. This discrepancy can be attributed to the simplicity of our approaches. In our model, we chose CNN as the feature extraction method, which is known for its rapid convergence to handle local features, especially compared to ViTs. This choice enabled our method to converge quickly in just 50 epochs. However, the lack of deeper CNNs or ViTs hinders accurate reconstruction of image details, resulting in lower SSIM values. This is one of the topics we will focus on in our future work. 

In summary, our approach seamlessly integrates traditional methods into deep learning paradigms, employs a straightforward structure to attain SOTA results, and highlights the synergy between traditional wisdom and modern deep learning techniques.

\begin{table}[ht]
    \centering
    \caption{Experimental results of image quality metrics. To evaluate the performance of the method proposed in this paper, we extracted the relevant comparison data from \cite{guo2023survey} and show them in this table. For each metric, the top three data points are shown in bold and underlined, bold only, and underlined only, respectively.}
    \begin{tabular}{lccc|cc}
        \toprule
        & \multicolumn{3}{c}{LOLv1} & \multicolumn{2}{c}{LOLv2} \\
        \midrule
        & PSNR$\uparrow$ & SSIM$\uparrow$ & LPIPS$\downarrow$ & PSNR$\uparrow$ & SSIM$\uparrow$  \\
        \midrule
        LIME~\cite{guo2016lime}            & 16.67 & 0.56 & 0.35 & 15.24 & 0.47 \\
        LightenNet~\cite{li2018lightennet}& 10.53 & 0.44 & 0.39 & - & - \\
        RetinexNet~\cite{Wei_2018_BMVC}      & 16.77 & 0.56 & 0.47 & 18.37 & 0.72 \\
        MBLLEN~\cite{Lv_2018_BMVC}          & 17.90 & {0.72} & \underline{0.25} & 18.00 & 0.72 \\
        EnGAN~\cite{jiang2021enlightengan}    & 17.48 & 0.65 & 0.32 & 18.23 & 0.62 \\
        KinD~\cite{zhang2019kindling}& \underline{20.87} & \textbf{0.79} & \textbf{\underline{0.17}} & 19.74 & \textbf{0.76} \\
        RUAS~\cite{liu2021ruas}& 18.23 & 0.72 & 0.35 & 18.37 & 0.72 \\
        IAT~\cite{Cui_2022_BMVC}& \textbf{\underline{23.38}} & \underline{\textbf{0.80}} & \textbf{0.21} & \textbf{\underline{23.50}} & \textbf{\underline{0.82}} \\
        \midrule
        CPGA-Net & \textbf{20.94} & \underline{0.75} & 0.26 & \textbf{20.79}& \textbf{0.76} \\
        CPGA-Net (DGF) & {20.31} & 0.70 & 0.29 & \underline{20.74} & \underline{0.73} \\
        \bottomrule
    \end{tabular}
    \label{tab:performance_metrics}
\end{table}

\begin{table}[ht]
    \centering
    \caption{Experiment results of efficiency. To evaluate the performance of the method proposed in this paper, we extracted the relevant comparison data from \cite{guo2023survey} and showed them in this table. For each metric, the top three data points are shown in bold and underlined, bold only, and underlined only, respectively. Also, we mark the top three methods with the best PSNR values in this table with * to provide a more comprehensive comparison between efficiency and image quality.}
    \label{tab:efficiency_results}
    \begin{tabular}{lccc}
        \toprule
        & \string#Param. (M)$\downarrow$ & FLOPs (G)$\downarrow$ & Time (s)$\downarrow$ \\
        \midrule
        LightenNet~\cite{li2018lightennet} & 0.028& - & - \\
        RetinexNet~\cite{Wei_2018_BMVC} & 0.555 & 587.47 & 0.120 \\
        MBLLEN~\cite{Lv_2018_BMVC} & 0.450 & 301.12 & 13.995 \\
        KinD*~\cite{zhang2019kindling} & 8.160 & 574.95 & 0.148 \\
        RUAS~\cite{liu2021ruas} & \textbf{\underline{0.003}} & \underline{\textbf{0.28}} & \textbf{0.006} \\
        IAT*~\cite{Cui_2022_BMVC} & 0.091 & \underline{{5.28}} &\underline{\textbf{0.003}} \\
        \midrule
        CPGA-Net* & \underline{0.025} & 6.03 & \underline{0.028} \\
        CPGA-Net (DGF)  & \textbf{0.018} & \textbf{1.10} & \textbf{0.006} \\
        \bottomrule
    \end{tabular}
\end{table}

\begin{table}[ht]
    \centering
    \caption{Experimental results using the NIQE metric on unpaired datasets. To evaluate the performance of the method proposed in this paper, we extracted relevant comparison data from \cite{zhang2019kindling} and show them in this table. For each metric, the top three data points are shown in bold and underlined, bold only, and underlined only, respectively.}
    \label{tab:niqe_results}
    \begin{tabular}{lcccc}
        \toprule
        & \multicolumn{3}{c}{NIQE$\downarrow$} \\
        \midrule
        & \textbf{MEF}~\cite{Ma_2015_TIP} & \textbf{LIME}~\cite{guo2016lime} & \textbf{NPE}~\cite{Wang_2013_TIP} & \textbf{Avg} \\
        \midrule
        Low-light Image & 4.265 & 4.438 & 4.319 & 4.341 \\
        Dong~\cite{dong2010fast} & 4.109 & 4.052 & 4.126 & 4.096 \\
        SRIE~\cite{fu2016weighted} & \textbf{3.475} & 3.788 & 4.113 & 3.792 \\
        MSR~\cite{Rahman_1996_ICIP} & 3.610 & \underline{3.764} & 4.366 & 3.913 \\
        NPE~\cite{Wang_2013_TIP} & \underline{3.524} & 3.905 & 3.953 & 3.794 \\
        LIME~\cite{guo2016lime} & 3.720 & 4.155 & 4.268 & 4.048 \\
        RetinexNet~\cite{Wei_2018_BMVC} & 4.149 & 4.420 & 4.485 & 4.351 \\
        EnGAN~\cite{jiang2021enlightengan} & \textbf{\underline{3.232}} & 3.719 & 4.113 & \textbf{3.688} \\
        KinD~\cite{zhang2019kindling} & 3.883 & \underline{\textbf{3.343}} & \underline{3.724} & \underline{\textbf{3.650}} \\
        RUAS~\cite{liu2021ruas} & 4.140 & 4.290 & 4.871 & 4.434 \\
        \midrule
        CPGA-Net  & 3.870 & \textbf{3.707} & \textbf{3.548} & \underline{3.708} \\
        CPGA-Net (DGF)  & 3.827 & {3.834} & \underline{\textbf{3.498}} & 3.720 \\
        \bottomrule
    \end{tabular}
\end{table}

\subsection{Ablation study}
The results of the ablation study are shown in Tables 4 and 5 and Figs. 7 and 8, which include the use of each module. First, the results indicate that using the channel prior yields performance comparable to using RGB information. However, upon incorporating the global branch, it becomes evident that the channel prior outperforms RGB in enhancing the results. This observation suggests that the channel prior is good at extracting crucial features that are essential for the accurate estimation of transmittance $t$.

Additionally, we explore exclusively utilizing $\widetilde{A}$-estimation, examining the significance of employing the reversed atmospheric scattering model. Our results provide clear insights into the effectiveness of integrating $t$-estimation into the process. Notably, the neural network achieves commendable results by including $t$-estimation, underscoring its importance. Furthermore, it becomes evident that $\widetilde{A}$-estimation employing a residual structure can extract and utilize more information from the input image and outperform other approaches. This observation aligns with prevailing trends in the field, highlighting the advantage of utilizing residual structures for image enhancement as well as the effectiveness of employing specialized and deeper network architectures.

Due to the global branch of our study, we discuss the efficiency of the IAAF and gamma estimation. Gamma estimation does increase PSNR by about 0.5 dB; however, SSIM decreases slightly, showing that global correction has a positive effect on contrast but no effect on structural details.

Additionally, we observed that the gamma component does not yield the expected results, deviating from the results in IAT \cite{Cui_2022_BMVC}. This divergence may stem from differences in scale and model characteristics. The local branch of IAT primarily comprises a specialized ViT structure, which is in sharp contrast to our model that utilizes CNN and channel priors. This discrepancy highlights that the combination of CNN and exponential domain shifting in training does not yield identical results to transformer-based methods. However, IAAF effectively addresses this issue by coordinating the objectives of local and global branches. Also, it is achieved by directly connecting them to the output, as opposed to letting the model learn directly and reducing the step-by-step processing of back-propagation complexity.

Regarding the selection of the gamma estimation module, integrating ResCBAM remarkably boosted PSNR by 0.6 dB compared to the plain module of ResBlock, although the SSIM values dropped slightly. This indicates its suitability for fitting gamma estimation within the global branch, thus improving the performance of the overall model.

Furthermore, our comprehensive analysis includes an extensive ablation study, as depicted in Fig. \ref{fig:ablationloss} and Table \ref{tab:ablation_loss}. The results of this study provide compelling evidence for the pivotal role of the perceptual loss in our model's performance, it provides a substantial 0.16 dB improvement in PSNR and a 1\% reduction in LPIPS, indicating that a significant enhancement in visual fidelity for both quantitative and perceptual aspects of image quality within our approach. In addition, self-supervised training improves performance with a 0.94 dB increase in PSNR, a 4\% improvement in SSIM, and a 2\% reduction in LPIPS. For CPGA-Net (DGF), the results indicate that knowledge distillation significantly improves overall performance, surpassing the performance without knowledge distillation and approaching the original performance achieved without utilizing the DGF module. This underscores the positive influence of knowledge distillation on the model's performance and validates the interpretability of our model with knowledge distillation, paving the way for further extensions and applications. Both self-supervised training and knowledge distillation demonstrate the powerful capabilities of pretraining for deep learning image enhancement.

\begin{table*}[ht]
    \centering
    \caption{Performance Analysis of network architecture on the LOLv1 dataset. CP (Channel Prior) encompasses the dark/bright channel prior and the Y channel from YCbCr. For $\widetilde{A}\text{-est}$ and $\gamma$-est, Conv, RBlk, and RCBAM represent CNN layer, ResBlock, and ResCBAM, respectively. IAAF stands for Intersection-Aware Adaptive Fusion Module for fusion strategy.The experiments in this table only use L1 loss and the uniform number of 16 channels to ensure fairness for all combinations. For each metric, the top three data points are shown in bold and underlined, bold only, and underlined only, respectively.}
    \label{tab:ablation}
    \begin{tabular}{lccccccc|ccc}
        \toprule
        & \multicolumn{2}{c}{$t$-est}  & \multicolumn{2}{c}{$\widetilde{A}\text{-est}$}& \multicolumn{2}{c}{$\gamma$-est}  &Fusion & \multicolumn{3}{c}{Image Quality Metrics} \\
        \midrule
        & RGB & CP & Conv & RBlk & RBlk & RCBAM & IAAF & PSNR & SSIM  & LPIPS \\
        \midrule
        (a) & & & \checkmark& & & & & 17.10 & 0.70 & 0.41 \\
        (b) & & & & \checkmark & & & & 17.64 & 0.72 & 0.40 \\
        (c) & \checkmark & & \checkmark & & & & & 18.22 & \textbf{0.74} & 0.33 \\
        (d) & & \checkmark & \checkmark & & & & & 18.24 & \underline{0.73} & 0.35 \\
        (e) & \checkmark & & & \checkmark & & & & 18.56 & \textbf{\underline{0.75}} & 0.34 \\
        (f) & & \checkmark & &\checkmark & & & & 18.59 & \underline{0.73} & \underline{0.32} \\
        (g) & & \checkmark & &\checkmark & \checkmark & & & 17.83 & 0.72 & 0.42 \\
        (h) & \checkmark & & & \checkmark & \checkmark & & \checkmark & \underline{19.04} & 0.71 & 0.34 \\
        (i) & & \checkmark & & \checkmark & \checkmark & & \checkmark & \textbf{19.24} & \underline{0.73} & \underline{\textbf{0.27}} \\
        (j) & & \checkmark & & \checkmark & & \checkmark & \checkmark & \textbf{\underline{19.84}} & 0.72 & \textbf{0.29} \\
        \bottomrule
    \end{tabular}
\end{table*}

\begin{table}[ht]
    \centering
    \caption{Performance Analysis of Loss Functions on the LOLv1 dataset. Pre-tr. denotes as pretrained weight for fine-tuning, and the units of \#Params. and FLOPs are M and G, respectively. SST stands for self-supervised training. (a) and (i) are the low-light image and normal-light image, respectively. (b) to (d) correspond to CPGA-Net with 16 channels of Resblocks, while (e) to (h) represent CPGA-Net with 8 channels of Resblocks. Specifically, (g) and (h) are instances of the CPGA-Net (DGF) structure with varying pretrained weights used for fine-tuning. The selected model of CPGA-Net and CPGA-Net (DGF) are (d) and (h), respectively. For each image quality metric, the top three data points of the experiments are shown in bold and underlined, bold only, and underlined only, respectively.}
    \label{tab:ablation_loss}
    \begin{tabular}{l|cccc|ccc|cc}
        \toprule
        & $L_1$ & $L_{per}$ & $L_{KD}$ & Pre-tr. & PSNR & SSIM & LPIPS&\#Param. & FLOPs \\
        \midrule
        (a) &  &  &  &  & 7.77 & 0.19 & 0.56 & - & - \\
        \midrule
        (b) & $\checkmark$ &  &  &  & 19.84 & {\textbf{0.72}} & \underline{0.29} & 0.025 & 6.03 \\
        (c) & $\checkmark$ & $\checkmark$ &  &  & 20.00 & \underline{0.71} & \textbf{0.28} & 0.025 & 6.03 \\
        (d) & $\checkmark$ & $\checkmark$ &  & SST & \textbf{\underline{20.94}} & \underline{\textbf{0.75}} & \underline{\textbf{0.26}} & 0.025 & 6.03 \\
        \midrule
        (e) & $\checkmark$ & $\checkmark$ &  &  & \underline{20.31} & {0.70} & \textbf{0.28} & 0.018 & 4.36 \\
        (f) & $\checkmark$ & $\checkmark$ & $\checkmark$ & (e) & {\textbf{20.75}} & 0.68 & \underline{\textbf{0.26}} & 0.018 & 4.36 \\
        (g) & $\checkmark$ & $\checkmark$ &  & (e) & 19.79 & {0.70} & 0.30 & 0.018 & 1.10 \\
        (h) & $\checkmark$ & $\checkmark$ &  & (f) & \underline{20.31} & {0.70} & \underline{0.29} & 0.018 & 1.10 \\
        \midrule
        (i) &  &  &  &  & $\infty$ & 1.00 & 0.00 & - & - \\
        \bottomrule
    \end{tabular}
\end{table}

\begin{figure}[ht]
    \centering
    \includegraphics[width=\textwidth]{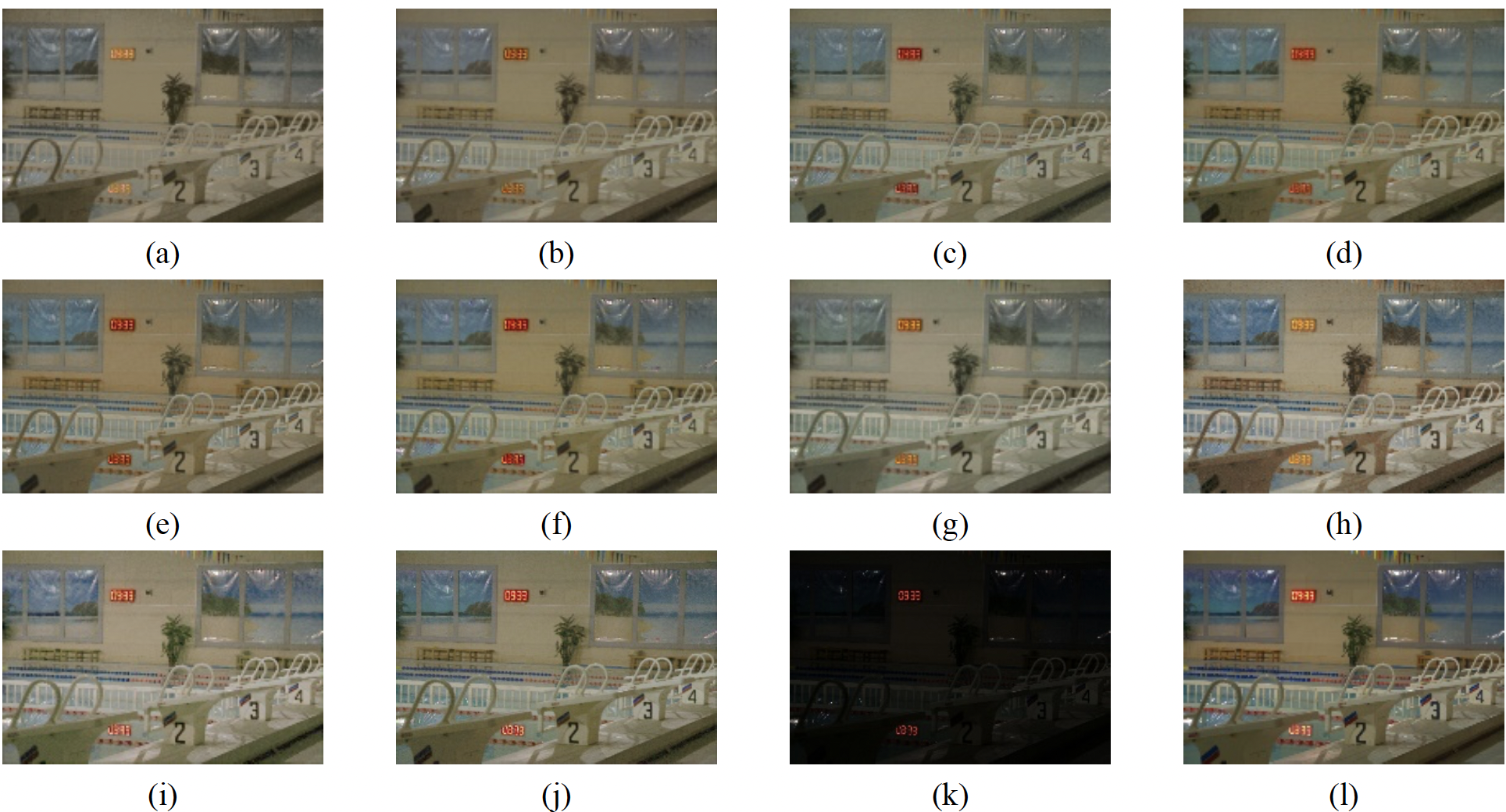}
    \caption{Visualization results of the network architecture ablation study. (a) to (j) follow the order of Table \ref{tab:ablation}, (k) represents the low-light image, and (l) represents the normal-light image.
}
    \label{fig:ablation}
\end{figure}

\begin{figure}[ht]
    \centering
    \includegraphics[width=\textwidth]{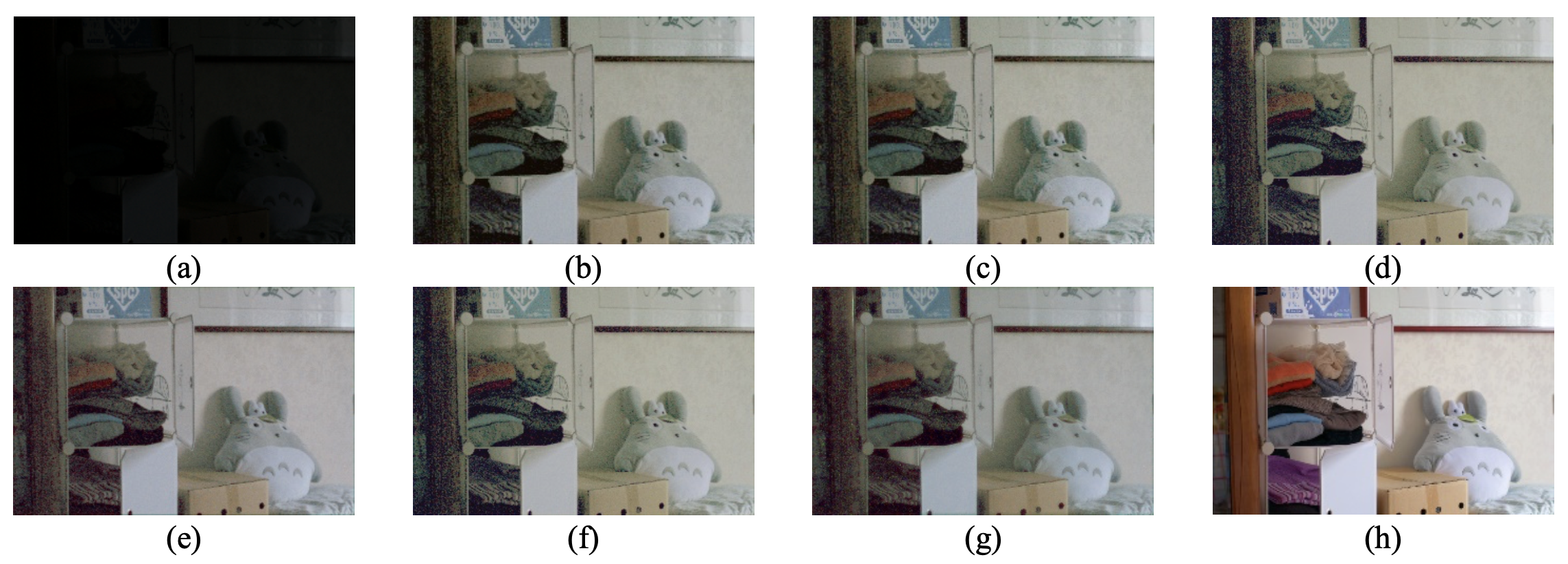}
    \caption{Visualization results of loss functions ablation study. (a) to (h) follow the order of Table \ref{tab:ablation_loss}.
}
    \label{fig:ablationloss}
\end{figure}

\subsection{Interpretability of CPGA-Net} \label{section: interpretability}
\begin{figure*}[ht]
    \centering
    \includegraphics[width=\textwidth]{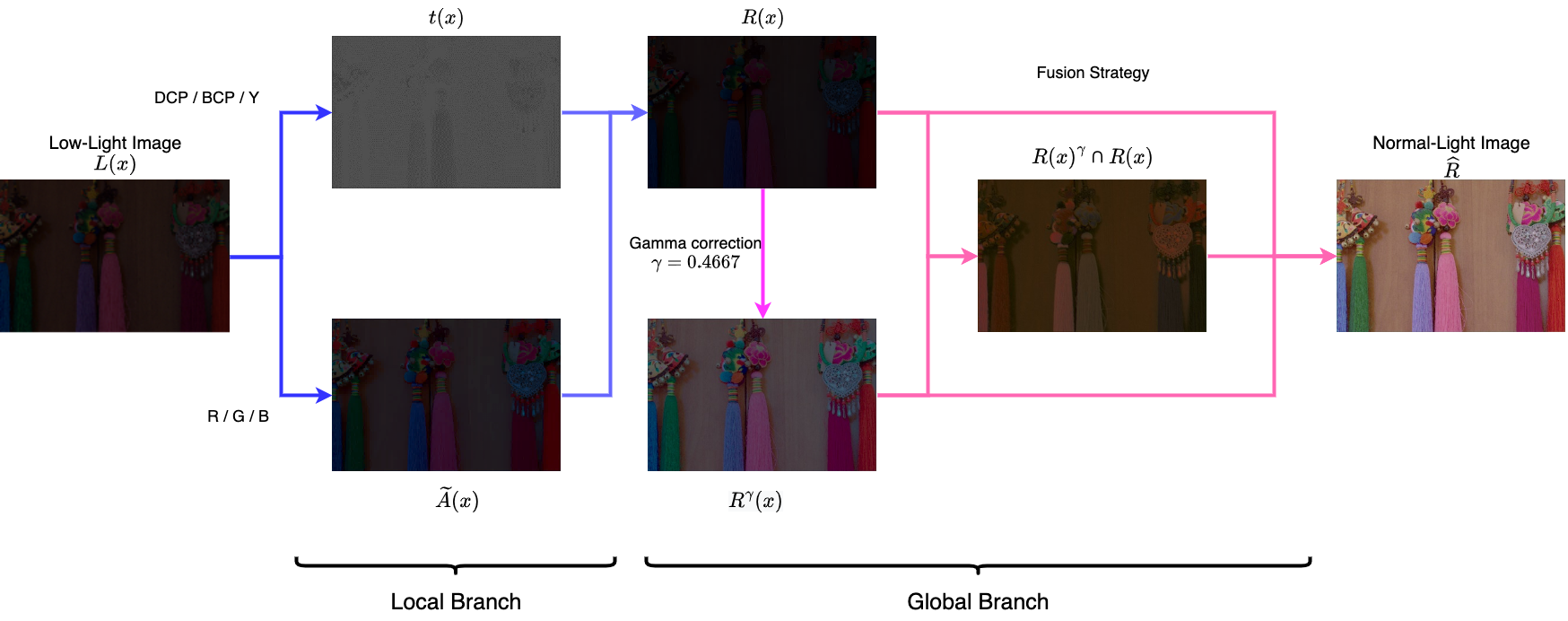}
    \caption{Visualization results of feature maps of CPGA-Net. For the visibility of observing $t$ and $\widetilde{A}$ of low-light images, we adjusted them with a 40\% brightness enhancement to make their features more apparent. For the intersection, we perform the absolute calculation to normalize the value to positive for visualization.
}
    \label{fig:interpretability}
\end{figure*}

The feature maps of CPGA-Net are shown in Fig. \ref{fig:interpretability}. As we know, the neural network is a black box that can not easily be understood by intuitive knowledge. However, our approach is fully designed using the theoretical formulations of ATSM, gamma correction, and fusion module based on set theory, and retains the features in three dimensions for the enhancement of RGB three channels output. We can look further to check out how various modules work in the neural network and connect the “black box” with the theory. Note that we only supervised our model with L1 loss and perceptual loss, there is no further supervision due to each module.

First of all, we can see that the lightness adjustment process is mostly been proceeded in the global branch instead of the local branch, and the local branch keeps in the low light conditions. It means that the process is split into two parts: information loss reconstructed by the local branch, and undesirable brightness shifts enhanced by the global branch. In our architecture, the $R$ in our model is not restricted and constrained by the $R$ from the theory.

For the local branch, we know that $\widetilde{A}$ is the fundamental backbone of restoring and keeping the color information calibrated to the correct tone. Compared with the results from Table~\ref{tab:ablation}, the ablation study of using the deeper module for $\widetilde{A}$-estimation, it is clear that its stronger capacity is one of the keys to improving the performance. $t$-estimation has the role of adjusting the balance of the contrast. In the feature map, we can see that the dim parts of $t$ usually occur in brighter and stronger color differences for structural and contrast enhancement. Overall, both parts have their roles of restoring the image and calibrating it correctly, we can know it from $R^\gamma$ with a correct color tone but darker features.

For the IAAF, the color tone of the intersection is complementary to the target tone and contains most of the characteristics of the output image, matching with the assumption of the intersection between two images. Note that it not only represents the similar regions of $R$ and $R^\gamma$, but it also stands for the connection of both for adjusting the brightness shift and unseen defects using the subtraction operator.

Last but not least, those observations are based on the results and the conjectures based on theoretical analysis, it is not a fixed phenomenon for different conditions or different combination rules. The roles of each module may vary due to the connection methods, properties of the input feature, and so on. However, it is still a big step to understand how neural networks know and the connection between deep learning and theoretical algorithms.

\section{Conclusion}
In this paper, we introduce CPGA-Net, a novel deep learning-based LLIE approach that incorporates traditional techniques such as channel priors and gamma correction. Importantly, CPGA-Net achieves this goal by employing fewer parameters than most existing methods. We demonstrate that CPGA-Net delivers image quality performance comparable to state-of-the-art methods, underscoring its effectiveness. This work represents a substantial contribution to the field of LLIE, uniting the fields of traditional learning and deep learning within a simple yet effective architectural framework with only essential feature extraction.

Our work also shows that the future of LLIE does not rely solely on large and complex deep learning architectures but strives to improve efficiency and accessibility, making these methods applicable to a wider range of applications beyond high-performance computing (HPC) environments. Our future research may focus on identifying detailed features of the architecture to cope with low performance, enhance its capability to handle various low-light conditions, and translate these methods into practical applications, such as mobile phones and embedded devices, bringing these advancements into the real world.









\bibliographystyle{ws-ijprai}
\bibliography{reference}




\end{document}